\documentclass[11pt,letterpaper]{article}
\usepackage{graphicx}
\usepackage{psfrag}
\usepackage{color}
\usepackage{amsmath}
\usepackage{amssymb}
\usepackage{lineno}
\usepackage{multirow}   
\usepackage[square,numbers,sort&compress]{natbib} 
\newcommand{\pt}[2]{\frac{\partial{#1}}{\partial{#2}}}
\newcommand{\ptn}[3]{\frac{\partial^{#3}{#1}}{\partial{#2}^{#3}}}
\newcommand{\dv}[2]{\frac{d #1}{d #2}}

\newcommand{\bfrac}[2]{\left(\frac{#1}{#2}\right)}

\newcommand{\kon}{k_{\textrm{on}}}
\newcommand{\koff}{k_{\textrm{off}}}

\newcommand{\dd}{\mbox{d}}
\newcommand{\mutypp}{\mbox{MutY}^{2+}}
\newcommand{\mutyppp}{\mbox{MutY}^{3+}}
\begin{document}

\title{Accelerated search kinetics mediated by redox reactions of DNA repair enzymes}

\author{Pak-Wing Fok\thanks{Dept. of Biomathematics, UCLA, Los Angeles, CA 90095-1766}\\
	Applied and Computational Mathematics \\
                             California Institute of Technology, \\
                            Pasadena,  CA 91125 and \\
                              Dept. of Biomathematics\\
	University of California, Los Angeles, CA 90095-1766
	\and Tom Chou\thanks{Dept. of Biomathematics, UCLA, Los Angeles, CA 90095-1766, 
                             Tel: 310-206-2787} \\
	Depts. of Biomathematics and Mathematics, \\
	University of California, Los Angeles, CA}

\date{}

\pagestyle{myheadings}
\markright{Accelerated Search of DNA Repair Enzymes}
\maketitle


\abstract{A Charge Transport (CT) mechanism has been proposed in
several papers ({\it e.g.}, Yavin \textit{et al.} PNAS \textbf{102}
3546 (2005)) to explain the localization of Base Excision Repair (BER)
enzymes to lesions on DNA.  The CT mechanism relies on redox reactions
of iron-sulfur cofactors that modify the enzyme's binding affinity.
These redox reactions are mediated by the DNA strand and involve the
exchange of electrons between BER enzymes along DNA.
We propose a mathematical model that incorporates enzyme
binding/unbinding, electron transport, and enzyme diffusion along DNA.
Analysis of our model within a range of parameter values suggests that
the redox reactions can increase desorption of BER enzymes not already
bound to their targets, allowing the enzymes to be recycled, thus
accelerating the overall search process. This acceleration mechanism
is most effective when enzyme copy numbers and enzyme diffusivity
along the DNA are small.  Under such conditions, we find that CT BER
enzymes find their targets more quickly than simple ``passive''
enzymes that simply attach to the DNA without desorbing.

%

}

\baselineskip=12pt
\section{Introduction}


The genomes of all living things can be damaged by ionizing radiation
and oxidative stress. These factors can cause mismatches in the DNA
strand resulting in localized lesions. The role of Base Excision
Repair (BER) enzymes is to locate and remove these lesions. If the
lesions are allowed to persist, they can give rise to mutations and
ultimately diseases such as cancer. 

\begin{figure}[h]
\begin{center}
\includegraphics[width=4in]{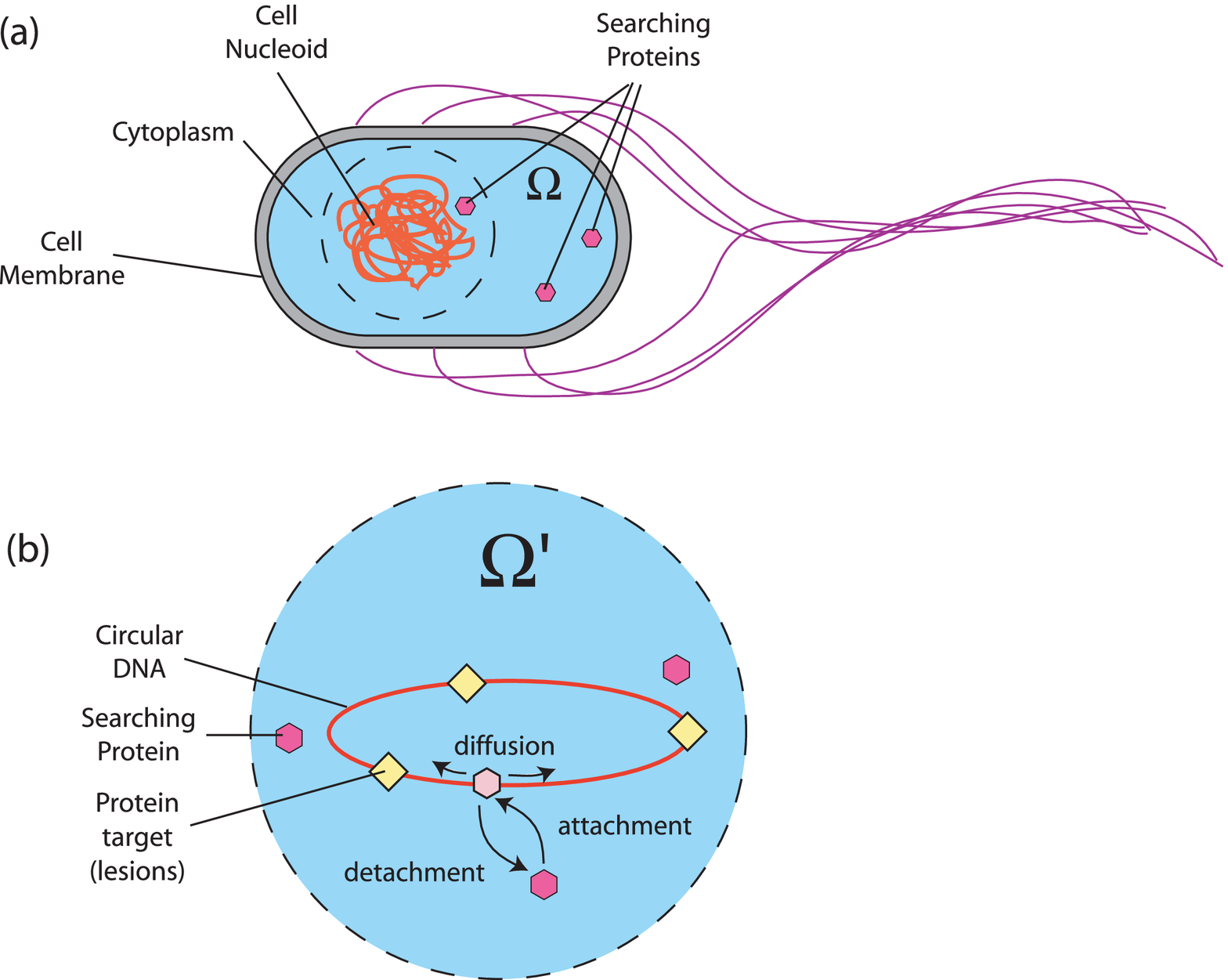}
\end{center}
\caption{}
%
\label{fig1}
\end{figure}

The localization of BER enzymes to lesions is physically related to
the binding of transcription factors to promoter regions that regulate
gene expression.  In 1970, experiments by Riggs \textit{et al.}
\cite{Riggs1,Riggs2} showed that the association rate of the LacI
repressor to its operator is about $100$ times faster than the maximum
rate predicted by Debye-Smoluchowski theory. This theory assumes that
LacI is transported to its target on DNA via 3D diffusion. To explain
the experimental observations, the theory was modified to account for
facilitated diffusion \cite{Berg81,Winter89,vonHippelBerg89}.  In this
process, the LacI repressor can spend part of its time attached to the
DNA and perform a 1D random walk before detaching and diffusing in 3D
again (see Fig. \ref{fig1}). Provided the protein spends about half
its time on the DNA and half its time in solution \emph{and} the
diffusivities in 1D and 3D are comparable, the predicted search time
can be reduced by as much as $100$ fold \cite{SlutskyMirny}. However,
these conditions are very restrictive as the protein can spend up to
99.99$\%$ of its time associated to the DNA \cite{WunderlichMirny} and
the diffusion constant along DNA (in 1D) is in general much smaller
than the one in the cytoplasm (in 3D) \cite{Wang}. Therefore, many
modifications of the basic facilitated diffusion theory have been
proposed, including intersegmental transfers \cite{kafri}, the effect
of DNA conformation \cite{Grosberg}, directed sliding \cite{Loverdo},
and finite protein concentration \cite{Cherstvy}.



A series of recent papers
\cite{BARTON,BARTON2,Boal} have revealed a special kind of long-ranged
interaction for certain BER enzymes based on charge transport (CT)
along DNA.  MutY, a type of DNA glycosylase, contains a
[4Fe-4S]$^{2+}$ cluster which is very sensitive to changes in its
environment. Specifically, its redox potential is modified depending
on whether it is in a polar environment (when the enzyme is in
solution) or in a more hydrophobic one (when the enzyme is attached to
DNA).  In solution, the [4Fe-4S]$^{2+}$ cluster is resistant to
oxidation.  However, when attached to DNA, the cluster is more easily
oxidized through the reaction [4Fe-4S]$^{2+}$ $\to$ [4Fe-4S]$^{3+}$ +
$e^-$. Furthermore, the $3+$ form has a binding affinity about
$10,000$ times greater than the $2+$ form \cite{Boal}.

\begin{figure}
\begin{center}
\includegraphics[width=4in]{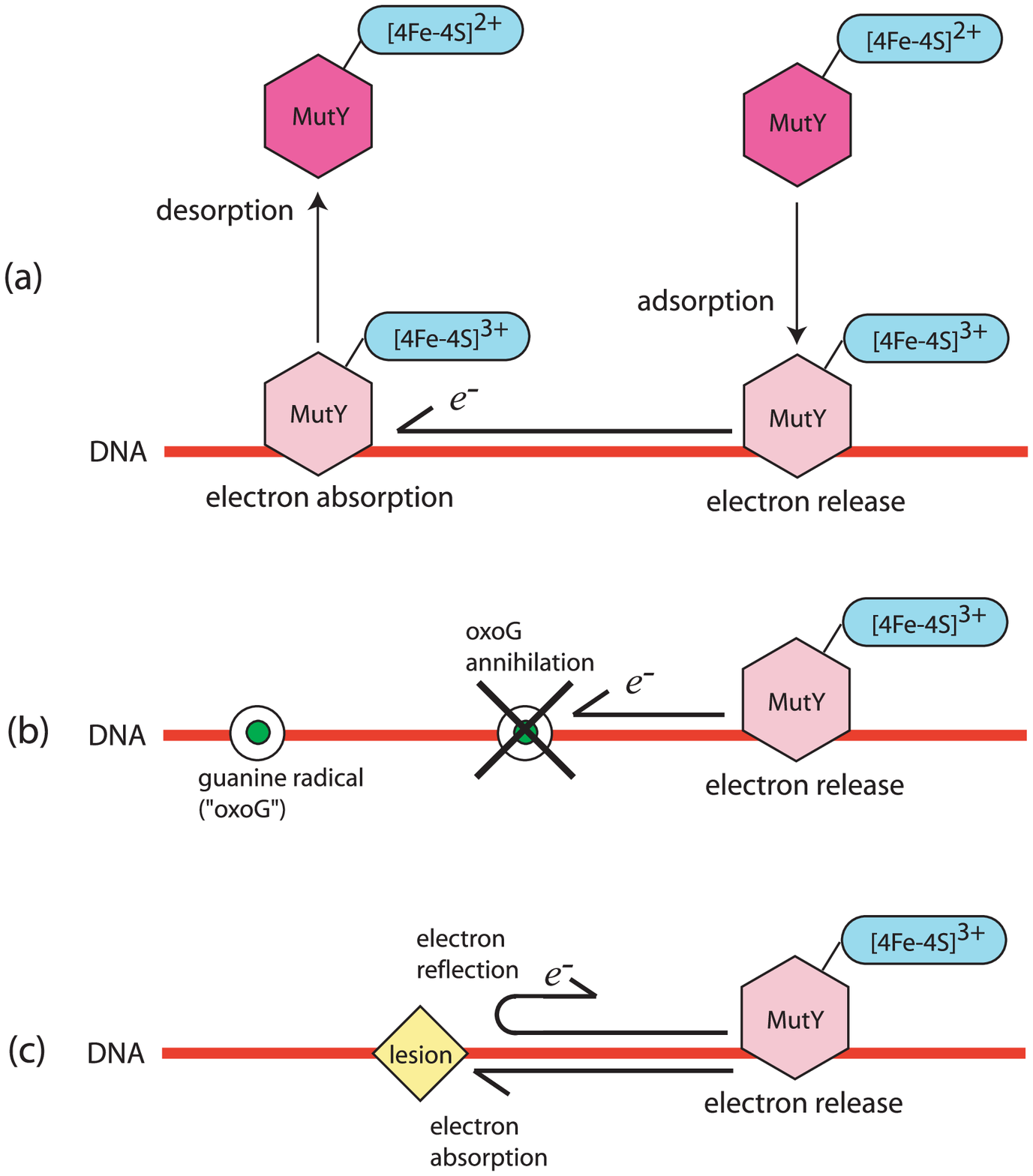}
\end{center}
\caption{}
%
%
\label{fig2}
\end{figure}

A model for the ``scanning'' of BER enzymes along DNA, aided by CT,
was proposed in \cite{BARTON,BARTON2,Boal}, and is depicted in
Fig.~\ref{fig2}.  When a BER enzyme adsorbs to DNA, it oxidizes and
releases an electron along the strand (see Fig. \ref{fig2}(a)).
Distal enzymes, already adsorbed onto the DNA can absorb these
electrons, become reduced and desorb. Hence, binding and unbinding of
enzymes are associated with oxidation and reduction of their iron-sulfur
clusters.  CT along DNA can be disrupted by the presence of defects
that affect electron transport. For example, guanine radicals
(``oxoGs''), formed under oxidative stress, can absorb electrons: see
Fig. \ref{fig2}(b).  By acting as sites of reduction, they promote the
adsorption of BER enzymes \cite{BARTON,FokChou}.  Once the radical has
absorbed an electron, it converts to a normal guanine base and no
longer participates in CT. However, ``permanent'' defects, or lesions,
can also exist on DNA which can absorb more than a single electron
(see Fig. \ref{fig2}(c)).
%
%
For example, oxoGs can erroneously pair with adenine bases when the
DNA replicates.  Such lesions may continuously absorb electrons with a
certain probability, or otherwise reflect them. In contrast to the
oxoG-cytosine case, the removal of oxoG-adenine lesions require MutY
to be present at the damaged site.

In this paper, we develop a model of CT-mediated BER enzyme kinetics
that includes enzyme diffusion along DNA, a binding rate that depends
on electron dynamics, and the effects of finite enzyme copy
number. Our key finding is that the proposed charge transfer mechanism
employed by BER enzymes accelerates their search for targets along DNA
in real finite enzyme copy number systems.
In the next section, we derive the governing
equations of enzyme kinetics. These equations are rendered
non-dimensional and key parameters are defined and estimated.  In
Section 3, we numerically solve our model equations under various
conditions and estimate the time for the binding of an enzyme to a
localized lesion.  We conclude with a discussion of our results in
Section 4.

\section{Mathematical model}
\subsection{Derivation of kinetic equations}

Consider the diffusion and adsorption-desorption kinetics of repair
enzymes in a bacterium such as \textit{E. coli}: see Fig. \ref{fig1}.
The chromosome in bacteria is circular but tightly coiled up into a
nucleoid that has an effective volume of about $8 \times 10^7
~\mbox{nm}^3$. If a repair enzyme is associated with the DNA strand,
it can diffuse freely along the DNA to find lesions. These associated
enzymes can spontaneously desorb from the strand, but they can also
become oxidized, leading to tighter binding to the DNA.  If later on,
the enzyme is reduced, its association with the DNA weakens and it can
quickly dissociate from the DNA. Localized lesions prevent the passage
of electrons (released along the DNA by oxidation of associated repair
enzymes) by either reflecting or absorbing them.

\begin{figure}[htbp]
\begin{center}
\includegraphics[width=4in]{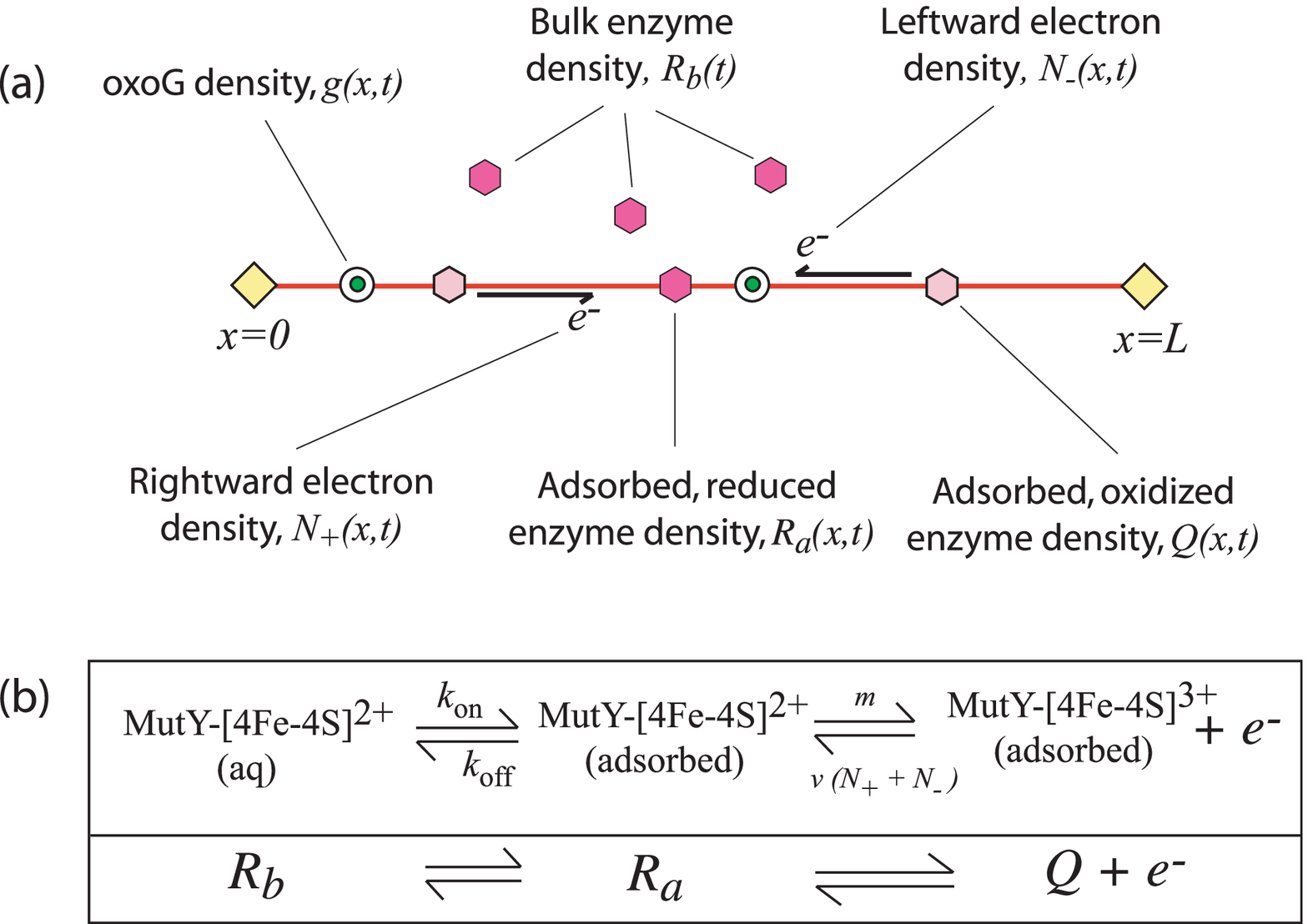}
\end{center}
\caption{}
%
\label{fig3}
\end{figure}

We write mass-action equations for the reactions occuring in
Fig. \ref{fig3}, coupled to equations that determine the electron
dynamics. We assume that the enzyme density in the bulk, $R_b(t)$
(where $t$ is time), is well mixed and has no spatial dependence.  The
density of DNA-adsorbed BER enzymes in the reduced and oxidized state
are denoted by $R_a(x,t)$ and $Q(x,t)$ respectively, where $0 \leq x
\leq L$ is the coordinate along the DNA and lesions are located at
$x=0$ and $x=L$.  The density of guanine radicals is $g(x,t)$ and the
density of rightward and leftward electrons is $N_+(x,t)$ and
$N_-(x,t)$. Note that $R_b(t)$ has units of inverse volume, while
$R_a(x,t)$, $Q(x,t)$, $N_{\pm}(x,t)$, and $g(x,t)$ carry units of
inverse length.  The governing equations corresponding to the
processes depicted in Figs. \ref{fig2} and \ref{fig3} are

\begin{eqnarray}
\pt{Q(x,t)}{t} &=& D_+ \ptn{Q}{x}{2} - v(N_++N_-)Q + m R_a, \label{eqn:CT1}\\
\pt{R_a(x,t)}{t} &=& D_- \ptn{R_a}{x}{2} + v(N_++N_-)Q - 
k_{\textrm{off}} R_a + k_{\textrm{on}} \bfrac{\Omega}{L} R_b - m R_a, \label{eqn:CT2}\\
\dv{R_b(t)}{t} &=& - k_{\textrm{on}} R_b + 
\frac{k_{\textrm{off}}}{\Omega} \int_0^L R_a dx, \label{eqn:CT3}\\
\pt{N_+(x,t)}{t} + v\pt{N_+(x,t)}{x} &=& f N_- - f N_+ - v N_+(Q + g) + 
\frac{m R_a}{2}, \label{eqn:CT4}\\
\pt{N_-(x,t)}{t} - v\pt{N_-(x,t)}{x} &=& -f N_- + fN_+ - v N_-(Q + g) + 
\frac{m R_a}{2}, \label{eqn:CT5} \\
\pt{g(x,t)}{t} &=& -v(N_++N_-) g. \label{eqn:CT6}
\end{eqnarray}
These equations must be solved subject to the boundary conditions
\begin{equation}
\begin{array}{cc}
N_+(0,t) = r N_-(0,t), & N_-(L,t) = r N_+(L,t), \\
Q(0,t) = Q(L,t) = 0, & R_a(0,t) = R_a(L,t) = 0,
\end{array}
\label{eqn:bcs}
\end{equation}
and initial conditions
\begin{equation}
\begin{array}{ccc}
Q(x,0) = 0, & R_a(x,0) = 0, & R_b(0) = n_0/\Omega, \\
N_+(x,0) = 0, & N_-(x,0) = 0, & g(x,0) = g_0/L.
\end{array}
\label{eqn:ics}
\end{equation}
In Eqs. \ref{eqn:CT1}-\ref{eqn:CT6}, $D_+$ is the diffusivity of
adsorbed MutY$^{3+}$ along the DNA, $D_-$ is the diffusivity of
adsorbed MutY$^{2+}$, $v$ is the speed of electrons along DNA, $m$ is
the electron release (oxidation) rate of adsorbed MutY$^{2+}$, $\koff$
is the intrinsic desorption rate of $\mutypp$, $\kon$ is the intrinsic
adsorption rate of $\mutypp$ to the DNA from solution, $\Omega$ is the
cell volume, $L$ is the arclength of the DNA, and $f$ is
the electron flip rate (see below).  In Eqs. \ref{eqn:bcs},
$r$ is the electron reflectivity of lesions, which we
describe in more detail later. In Eqs. \ref{eqn:ics}, $n_{0}$ is the copy number of
MutY, and $g_0$ is the initial number of guanine radicals on the DNA.
The definitions of all constants are summarized in Table
\ref{tab:constants}.

\begin{table}[htbp]
\begin{center}
\begin{tabular}{|c|c|c|c|}
\hline
Symbol & Definition & Typical value & Reference \\[12pt]
\hline \hline
$D_{\pm}$ & Diffusivity of adsorbed enzymes & $5 \times 10^6$ bp$^2$/s & \cite{XIE} \\[11pt]
$v$ & Electron velocity & $10^{10}$ bp/s & \cite{BARTON3}\\[11pt]
$f$ & Electron flip rate & $10^9 - 10^{10}$ s$^{-1}$ & $^1$\\[11pt]
$m$ & Electron release rate & $\sim 10^6$ s$^{-1}$ & \cite{Lin} \\[11pt]
$\Omega$ & Bacterium volume & $3.7 \times 10^8 \mbox{nm}^3$ & \cite{Woldringh}\\[11pt]
$L$ & Length of DNA & $5 \times 10^6$ bp & \\[11pt]
$\kon$ & MutY$^{2+}$ attachment rate & 2000 s$^{-1}$& \cite{FokChou} \\[11pt]
$\koff$ & MutY$^{2+}$ detachment rate & $7 \times 10^{-3}$ s$^{-1}$ & $^2$ \\[11pt]
$n_{0}$ & Copy number of MutY in \textit{E. coli} & 20-30 & \cite{Bai,Demple} \\[11pt]
$r$ & Electron reflectivity of lesions & 0~--~1 & - \\[11pt]
$g_0$ & Number of oxoGs on \textit{E. coli} DNA & $\sim 30$ & $^3$ \\[11pt]
\hline
\end{tabular}
\end{center}
\caption{Key constants for used for repair enzyme model Eqs. \ref{eqn:CT1}-
\ref{eqn:CT6} and conditions \ref{eqn:bcs},\ref{eqn:ics}.}
\label{tab:constants}
\end{table}

\footnotetext[1]{The mean free path of an electron
is estimated to be $\lambda \sim 1-10$ base pairs
and the flip rate approximated as $v/\lambda$.}


\footnotetext[2]{Estimated using the time taken for the restriction
endonuclease BsoBI to unbind from DNA \cite{Koch},
$t_{\textrm{off}}=$150s and taking $\koff = 1/t_{\textrm{off}}$. This
value of $t_{\textrm{off}}$ may not be an accurate value for the
unbinding time for MutY.}

\footnotetext[3]{Assumes that about 1 in 40,000 guanine bases are
oxoGs \cite{Helbock} and the length of \textit{E. coli} DNA is $L = 5
\times 10^6$ bp.}

\begin{table}[htbp]
\begin{center}
\begin{tabular}{|c|c|c|}
\hline
Parameter & Definition & Calculated Value \\[12pt]
\hline \hline
$\eta$ & $D_+/(\kon L^2)$ & \parbox[c]{1in}{$\lesssim 10^{-10}$} \\[12pt]
$U$ & $v/(\kon L)$ & \parbox[c]{1in}{$5$} \\[12pt]
$\sigma$ & $m/(m+\koff)$ & \parbox[c]{1in}{$\sim 1$} \\[12pt]
$F$ & $f/\kon$ &  $5 \times 10^5 - 5 \times 10^6$ \\[12pt]
\hline
\end{tabular}
\caption{Dimensionless parameters in Eqs. \ref{eqn:fQ}-\ref{eqn:fg}}
\label{tab:constants2}
\end{center}
\end{table}

We now give a brief justification of equations
\ref{eqn:CT1}-\ref{eqn:CT6} and conditions \ref{eqn:bcs} and
\ref{eqn:ics}.  The form of the first three equations can be
understood from Fig. \ref{fig3}(b) which summarizes the reactions
among the three species $R_b$, $R_a$, and $Q$.  Eq. \ref{eqn:CT1}
describes the time rate of change of adsorbed $\mutyppp$ 
due to oxidation of adsorbed $\mutypp$ ($+m R_a$) and reduction by
incoming electrons ($-v(N_+ + N_-)Q$).  The first term on the right
hand side represents diffusion along the DNA.
Eq. \ref{eqn:CT2} describes the evolution of adsorbed $\mutypp$ in
terms of the reduction of $\mutyppp$ ($+v(N_+ + N_-)Q$), spontaneous
desorption into solution ($-\koff R_a$), adsorption of aqueous
$\mutypp$ ($\kon (\Omega/L) R_b$) and oxidation into $\mutyppp$
($-mR_a$). Since $\mutypp$ binds to DNA less strongly than $\mutyppp$,
it is possible that $D_+$ is appreciably smaller than $D_-$.
Eq. \ref{eqn:CT3} is an equation for the concentration of $\mutypp$ in
solution which can decrease by enzymes binding to the DNA ($-\kon
R_b$) and increase by enzymes unbinding from the DNA (represented by
the integral term). Because we assume enzymes in the bulk solution are
well mixed, any increases in bulk concentration are due to an
integrated DNA-adsorbed density which does not distinguish between
enzymes that are released from different positions along the DNA, but
only sees the total number of enzymes that desorb.

Eqs. \ref{eqn:CT4} and \ref{eqn:CT5} describe the electron dynamics.
In our model, right and left-moving electrons (see
Fig. \ref{fig3}(a)) propagate along the DNA with speed $v$; this
process is represented by the two convective terms on each of the left
hand sides.  Also, electrons are lost when they are absorbed by
$\mutyppp$ or by guanine radicals, and produced when released by
adsorbed $\mutypp$. These processes are represented by the third and
fourth terms on the right hand side of Eqs.~\ref{eqn:CT4} and
\ref{eqn:CT5}, respectively. Finally, leftward and rightward electrons
can inter-convert \cite{FokChou} by scattering off inhomogeneities and
thermally induced conformational changes in the DNA
\cite{MDO0,MDO1}. This process is represented by the first and second
terms on the right hand side. The flip rate $f$ characterizes how
frequently a traveling electron changes direction. If $f$ is large,
the electron move diffusively, but if $f$ is small, it moves in a  more
ballistic manner.
Finally, Eq.~\ref{eqn:CT6} represents the evolution of the guanine
radical population. OxoGs are annihilated when they absorb electrons
as represented by the $-v(N_+ + N_-)g$ term. Radicals might also be
spontaneously generated and modeled by a source term on the right hand
side of Eq.~\ref{eqn:CT6}. In this paper, we neglect spontaneous oxoGs
generation.

Equations \ref{eqn:CT1}-\ref{eqn:ics} use a mean field approximation
that neglects stochastic fluctuations in enzyme, electron and guanine
number. The effect of noise in the system could be included through
the use of a chemical master equation \cite{Arkin_nature}; however,
generalizing the equation to account for spatial variations along the
DNA is beyond the scope of this paper \cite{Isaacson}. Nonetheless, we
expect our results for lesion targeting by enzymes will be
qualitatively accurate.

Lesions at $x=0$ and $x=L$ (see Fig. \ref{fig3}(a)) define the domain
of solution for Eqs.~\ref{eqn:CT1}-\ref{eqn:CT6} which are subject to
the boundary conditions \ref{eqn:bcs}.  This domain can represent a
circular DNA with circumference $L$ containing a single lesion. In the
first equation of \ref{eqn:bcs}, leftward traveling electrons are
converted to rightward traveling ones by the lesion that reflects
leftward electrons with probability $r$. If $r=0$, leftward electrons
are absorbed by the lesion. On the other hand, if $r=1$, the lesion is
fully reflective and the rightward and leftward electron densities are
equal. Similar considerations apply to the lesion at $x=L$.  Since we
will eventually use our mean-field mass action equations to estimate
the mean time for a repair enzyme to find a lesion, we assume that the
lesions are perfectly ``absorbing'' and set $Q=R_{a}=0$ at the lesion
positions.  Our simulations are performed on a domain with $g_0$ oxoG
radicals and a bulk solution that contains $n_{0}$ enzymes (see
Eqs.~\ref{eqn:ics}); hence the adsorbed oxoG density is $g_0/L$ and
the bulk concentration is $n_{0}/\Omega$.

\subsection{Model reduction and non-dimensionalization}
Before non-dimensionalizing Eqs.~\ref{eqn:CT1}-\ref{eqn:CT6} we can
make one important simplification. On the right hand side of
Eq.~\ref{eqn:CT2}, the sizes of the second, third, fourth and fifth
terms are approximately $v/L^2$, $\koff/L$, $\kon/L$ and $m/L$ (in
units of bp$^{-1}$ s$^{-1}$) respectively. Guided by Table
\ref{tab:constants}, we assume the term $m R_a$ dominates. Since the
oxidation rate is large, adsorbed $\mutypp$ quickly oxidizes into the
$3+$ form upon adsorption onto DNA. For times $t \gg 1/m$ and rates
$k_{\rm{off}}+m \gg v/L, k_{\rm{on}}$ Eq. \ref{eqn:CT2} gives
$R_a(x,t) \ll 1$ for all $0 \leq x \leq L$ and we can neglect spatial
gradients in $R_a$ as well as $\partial R_{a}/\partial t$. Therefore,
we approximate Eq.~\ref{eqn:CT2} with
\begin{equation}
R_a(x,t) \approx \frac{1}{m+\koff} \left(
v(N_+ + N_-)Q + \kon \bfrac{\Omega}{L} R_b
\right).
\label{RAAPPROX}
\end{equation}
Upon substitution of Eq.~\ref{RAAPPROX} into Eqs.~\ref{eqn:CT1},
\ref{eqn:CT4} and \ref{eqn:CT5}, we eliminate the equations for
$R_{a}$ and find a reduction analogous to one commonly used in
deriving the steady-state limit of Michaelis-Menten kinetics
\cite{Murray}.

We now non-dimensionalize our equations by measuring time in units of
$\kon^{-1}$, length in units of $L$, concentration of adsorbed species
in units of $1/L$ and concentration of bulk species in units of
$1/\Omega$. Our final set of reduced and nondimensionalized equations
that describe the transport and kinetics of MutY repair enzymes, right
and left-moving electrons, and guanine radicals is
\begin{eqnarray}
\pt{Q(x,t)}{t} &=&   - U(1-\sigma) (N_+ + N_-) Q  + \eta \ptn{Q}{x}{2} 
+ \sigma  R_b , \label{eqn:fQ} \\[13pt]
\dv{R_b(t)}{t} &=&  U(1-\sigma) \int_0^1 (N_+ + N_-) Q dx - \sigma  R_b, 
\label{eqn:fRb} \\[13pt]
\pt{N_+(x,t)}{t} + U \pt{N_+(x,t)}{x} &=& F(N_- - N_+) - 
g U N_+ + \frac{\sigma R_b}{2} - \left( 1-\frac{\sigma}{2} \right) U N_+ Q + \frac{\sigma}{2}~ 
U N_- Q,  \label{eqn:fNp} \\[13pt]
\pt{N_-(x,t)}{t} - U \pt{N_-(x,t)}{x} &=& -F(N_- - N_+) - 
g U N_- + \frac{\sigma R_b}{2} + \frac{\sigma}{2} U N_+ Q - 
\left( 1-\frac{\sigma}{2} \right) U N_- Q, \label{eqn:fNm} \\[13pt]
\pt{g(x,t)}{t} &=& -U(N_++N_-) g, \label{eqn:fg}
\end{eqnarray}
where we have defined the dimensionless quantities

\begin{equation}
\eta = \frac{D_{\pm}}{\kon L^2}, \quad U = \frac{v}{\kon L}, 
\quad F = \frac{f}{\kon},
\label{eqn:dimensionless_const}
\end{equation}
and
\begin{equation}
\sigma \equiv \frac{m}{m+\koff},
\label{eqn:sigma}
\end{equation} 
%
%
which can be estimated using Table \ref{tab:constants2}. As we discuss
later, the parameter $\sigma$ represents the effective binding rate in
terms of the competition between the electron release rate $m$ and the
desorption rate of DNA-bound MutY$^{2+}$ $\koff$, and lies between 0
and 1. The dimensionless boundary and initial conditions are
\begin{equation}
\begin{array}{ccc}
N_+(0,t) = r N_-(0,t), & N_-(1,t) = r N_+(1,t), & Q(0,t) = Q(1,t) = 0,
\end{array}
\end{equation}
and
\begin{equation}
\begin{array}{rl}
Q(x,0) = 0, & R_b(0) = n_0, \\
N_+(x,0) = N_-(x,0) = 0, & g(x,0) = g_0.
\end{array}
\end{equation}
Our model can approximate the case of ``infinite'' enzyme copy number
when the transport of bulk enzymes is diffusion-limited. Although most
of the enzymes cannot immediately adsorb onto the DNA as they are too
far away, we assume that a certain number, $R_b$, are in the vicinity
of the nucleoid, say within a volume $\Omega'$ (see Fig. \ref{fig1}),
and are able to directly engage in adsorption.  However, instead of
being depleted over time, $R_b$ is continuously replenished by far
enzymes that diffuse into $\Omega' \subset \Omega$ to keep $R_b$
fixed.  Therefore, to obtain the infinite copy number limit, we hold
$R_b$ constant in Eqs. \ref{eqn:fQ}, \ref{eqn:fNp} and \ref{eqn:fNm}
and Eq.~\ref{eqn:fRb} no longer applies. To summarize, we model the
infinite copy number case by holding $R_b$ constant. In the finite
copy number case, $R_b(t)$ is allowed to vary in time through
Eq. \ref{eqn:fRb}.  Finally, note that equations describing a simple
diffusing enzyme that does not undergo CT can be recovered from
Eqs. \ref{eqn:fQ}-\ref{eqn:fg} by setting $U=0$. In this case, the
equations for $Q(x,t)$ and $R_b(t)$ decouple from the rest.

\subsection{Repair enzyme binding affinity $\sigma$}
In Eqs. \ref{eqn:fQ}-\ref{eqn:fg}, the rate of creation of reduced,
adsorbed enzyme $R_a$ from reduced bulk enzyme $R_b$ is exactly $R_b$
since we measure time in units of $1/\kon$. However, the overall rate
of the compound reaction $R_b \rightleftharpoons R_a \rightarrow Q$ is
$\sigma R_b$. Consider a $\mutyppp$ that is adsorbed onto the DNA. If
it absorbs an incoming electron, it can either desorb into the bulk or
it can release an electron back along the DNA and remain oxidized. The
parameter $\sigma$ in Eq. \ref{eqn:sigma} is the probability of
electron release. When $\koff \gg m$, a $\mutyppp$ that absorbs an
electron will preferentially desorb ($R_a \rightarrow R_b$), but when
$\koff \ll m$ a $\mutyppp$ will simply release the electron it just
absorbed to stay adsorbed onto the DNA ($R_a \rightarrow Q$). These
limiting behaviors are realized by taking $\sigma \rightarrow 0$ and
$\sigma \rightarrow 1$ respectively.

If $\sigma \sim 0$, a bulk reduced enzyme that adsorbs onto the DNA
quickly desorbs back into the bulk, while if $\sigma \sim 1$,
$\mutypp$ on the DNA prefers to oxidize and stay adsorbed rather than
go into solution.
Once it is oxidized, any further electrons that are absorbed will be
re-emitted in a random direction. Hence the electron changes direction
with probability $1/2$ whenever it encounters an adsorbed $\mutyppp$:
when $\sigma=1$, the terms with prefactors $(1-\sigma/2)$ and
$\sigma/2$ in Eqs.  \ref{eqn:fNp} and \ref{eqn:fNm} add to the
$F(N_{-}- N_{+})$ terms to yield an effective flip rate of
$F+UQ/2$. The seeding of oxidized enzymes on the DNA increases the
effective electron flipping rate because these enzymes can absorb
electrons and immediately release them back along the DNA in the
direction they came from or in the direction they were going.

We end this section with the comment that the model for the CT redox
process in Fig. \ref{fig2} is not exactly equivalent
to the reaction scheme in Fig. \ref{fig3}(b).  In Fig
\ref{fig2}, a bulk MutY$^{2+}$ ($R_b$) adsorbs onto a DNA and
immediately oxidizes, releasing an electron along the DNA. DNA-bound
MutY$^{3+}$ ($Q$) remains adsorbed until it absorbs an incoming
electron, whereupon it reduces and immediately desorbs into the
bulk. For this model to hold, the reaction kinetics in
Fig. \ref{fig3}(b) must be non-Markovian. Specifically, consider the
intermediate quantity $R_a$ in Fig. \ref{fig3}(b). An $R_a$ enzyme
oxidizes to a $Q$ enzyme ($R_a \to Q$) only if it `remembered' that it
was originally created via a $R_b \rightarrow R_a$ reaction. Likewise,
an $R_a$ enzyme desorbs ($R_a \to R_b$) only if it `remembered' that
it was originally created through a $Q \rightarrow R_a$ reaction.

\section{Results and Discussion}

We now compute and analyze solutions to Eqs. \ref{eqn:fQ}-\ref{eqn:fg}
for the infinite and finite copy number cases.  The equations are
solved numerically using second order finite differences on a
non-uniform grid that clusters grid points near the boundaries and a
trapezoidal rule to approximate the integrals. MATLAB's stiff solver
$\texttt{ode15s}$ was used to integrate the equations in time.  In the
infinite case $R_b$ is held at the value $n_{0}$ and in the finite case,
$R_b(t)$ is included in the dynamics with initial condition $R_b(0) =
n_0$.  Furthermore in each case we consider the dynamics associated
with CT enzymes where $U >0$, and the dynamics associated with
``passive'', non-CT enzymes where $U=0$. Setting $U=0$ decouples the
equations for electron and guanine radical dynamics
(Eqs. \ref{eqn:fNp}-\ref{eqn:fg}) from the equation for $Q(x,t)$,
the density of DNA-bound enzymes (Eq. \ref{eqn:fQ}).

We shall explore the behavior of Eqs. \ref{eqn:fQ}-\ref{eqn:fg},
and the associated search times defined below, with respect to:
\begin{itemize}
\item $\sigma$, the effective binding affinity.  Generally we have $0
< \sigma < 1$. From the values of $m$ and $\koff$ in Table
\ref{tab:constants}, we have $\sigma \approx 10^{-8}$.  This value of
$\sigma$ renders the desorption term $-U(1-\sigma)(N_+ + N_-)Q$ in
Eq. \ref{eqn:fQ} insignificant, making the effect of CT
negligible. Therefore, a necessary requirement for an effective CT
mechanism is that $\sigma < 1$.  In our simulations for the MutY
system, we take $\sigma=0.9$ bearing in mind that the value of $\koff$
in Table \ref{tab:constants} is for BsoBI and not MutY.
\item $\eta$, the diffusivity of MutY$^{3+}$ along DNA.  The value in
Table \ref{tab:constants2} of $\eta = 10^{-10}$ is based on the
diffusive sliding of a human glycosylase, hOgg1, which has a
diffusivity of about $5 \times 10^6$ bp$^2$/s \cite{XIE}. However, this 
value may not necessarily be an accurate value for MutY. Therefore we
will explore a range of diffusivities $\eta$ near $10^{-10}$.
\item $g_0$, the initial guanine radical density: There are about 30
oxoGs at any given time on \textit{E. coli} DNA, but
this number depends on environmental conditions. Hence we
explore a range of values centered around $g_0 = 30$.
\item $r$, the lesion reflectivity: the interaction between electrons
and lesions depends on unknown molecular factors at the lesion and in
the bulk cytoplasm.  Hence, we explore a full range of $r$ values
between 0 and 1.
\item $F$, the electron flip rate: the precise dynamics of electrons
on DNA is a very complicated process; our estimate for $F$ in Table
\ref{tab:constants2} makes many simplifications and may not be
accurate. We will explore a range of $F$ values centered around
$10^5$.
\end{itemize}

Figure \ref{fig4} shows snapshots of adsorbed enzyme, guanine and
electron density profiles for a finite enzyme copy number ($n_{0}=30$)
system. The profiles are shown near the lesion at $x=0$ at times $t=2$
and $t=5$.  The electron density is generally smaller at the lesions
and larger in the middle of the domain, resulting in a larger enzyme
desorption rate away from lesions (the desorption rate in
Eq. \ref{eqn:fQ} is proportional to the total electron density $N_+ +
N_-$). Thus, the CT enzyme density is smaller than that for passive
enzymes away from lesions. The enhanced desorption of CT enzymes from
the interior continuously replenishes the number of enzymes in
solution so that $R_b(t)$ decreases less rapidly than for passive
enzymes. For intermediate times, the net deposition rate is larger for
CT enzymes, the enzyme density near the lesions is also larger
(Fig. \ref{fig4}(a)) and grows in time (Fig. \ref{fig4}(b)). For long
times, the density vanishes everywhere: this is the trivial steady
state solution to Eqs. \ref{eqn:fQ}-\ref{eqn:fg}.


Figure \ref{fig5} shows the DNA-bound enzyme density at $t=40$. 
In (a), there is a sharp spike in the enzyme density near
the lesion at $x=0$, but otherwise the enzyme density is
relatively small. Note that all densities are symmetric about $x=1/2$. 
In Eq. \ref{eqn:fQ}, CT enzymes desorb with a
rate proportional to the total electron density $N_+ + N_-$. As seen
in (a), this density is smallest at the
lesions. Therefore the enzyme density near $x=0$ and $x=1$ grows more
quickly compared to the interior density.  The inset shows a rapid
variation in $Q$ of about 600 within a boundary layer of width $\sim
10^{-3}$. Using a non-uniform grid that clusters the
mesh points near the boundaries, we are able to resolve these boundary
layers to calculate the flux of enzymes through the lesions.  In
Fig. \ref{fig5}(b), CT $(U=1)$ and passive $(U=0)$ enzyme densities
are compared when the copy number is finite.  The CT-enzyme density
has sharp maxima near the lesions, while the passive enzyme density
does not. Compared to the infinite copy number case, the size of the
maxima is smaller since the number of enzymes in the bulk (and hence
the deposition rate) decreases with time.  Because of the maxima, the
flux of enzymes into the lesion,
\begin{equation}
J(t) = \eta\left[{\partial Q(x,t)\over \partial x}
+{\partial R_{a}(x,t)\over 
\partial x} \right]_{x=0}
\label{eqn:J}
\end{equation}
is greater compared to the passive case.  Figure \ref{fig5}(c)
compares the current for CT $(U=1)$ and passive enzymes $(U=0)$ when
the copy number is infinite. The passive enzyme current is always
greater than the CT enzyme current because for a constant deposition
rate, any desorption reduces the number of enzymes on the DNA and the
flux of enzymes into the lesion.  Therefore, for infinite copy number
systems, search by passive enzymes will always be faster than CT
enzymes.
In contrast, when the bulk contains a finite number of enzymes,
Fig. \ref{fig5}(d) shows that the CT current is always greater than
the passive current.  This is due to free electrons on the DNA that
determine the local desorption rate.  In the CT mechanism, enzymes are
knocked off the DNA by incoming electrons and on average, \emph{desorb
from lesion-free portions of the DNA and re-adsorb at positions closer
to the lesion}. The result is that for intermediate times ($t \gtrsim
10$), the current experiences a second growth phase, a behavior that
is never seen for the passive case. Ultimately, we have $J(t) \to 0$
as $t \to \infty$ for both passive and CT enzymes, but CT ensures that
this behavior occurs at a much later time.


Next, we consider the typical time for the first enzyme to reach
a lesion.  Since the enzyme density is symmetric about $x=1/2$,
the total flux can be found by using twice the enzyme flux to one
lesion defined in Eq. \ref{eqn:J}.
The typical search time $\tau_{s}$ is then approximated by integrating
$2 J(t)$ until one enzyme has diffused into the lesion:
\begin{equation}
\int_0^{\tau_{s}} 2 J(t) dt \approx 1.
\label{TAUS}
\end{equation}
From solving the full set of equations \ref{eqn:CT1}-\ref{eqn:CT6}
numerically, we find that the gradients in $R_{a}(x,t)$ at the lesions
are negligible compared to those of $Q(x,t)$, verifying the validity
of eliminating $R_{a}$ and using $J(t)\approx \eta (\partial
Q(x,t)/\partial x)_{x=0}$ as the total enzyme current. In the mean
field limit, an alternative definition of the search time is $\tau_{s}
\approx \int_{0}^{\infty}t \exp\left[-\int_{0}^{t}J(t')\dd
t'\right]\dd t$. We have computed $\tau_{s}$ using this mean field
approximation and find negligible qualitative differences from
$\tau_{s}$ computed using Eq. \ref{TAUS}.

%
%
%
Figure \ref{fig6}(a) shows that the search times are extremely
sensitive to the initial number of oxoGs $g_0$. In particular, there
is a rapid increase in $\tau_{s}$ as $g_0$ increases past the enzyme
copy number $n_0 = 30$. The CT mechanism relies on the presence of
free electrons that cause enzymes to desorb from lesion-free portions
of the strand and re-adsorb near lesion sites, while oxoGs suppress CT
by absorbing free electrons. When $g_0 > n_0$, all enzymes from the
bulk adsorb onto the DNA and any released electrons are absorbed by
nearby guanine radicals.  Instead of participating in CT-mediated
redistribution and localization, the enzymes cannot desorb and must
rely on slow diffusive sliding along the DNA strand to find their
targets.
%
%
%
When $g_0 < n_0$, at least one enzyme is always in solution and is
transported through the cytoplasm. Since 3D transport is assumed to be
fast, the search time is correspondingly small.  Also in this plot,
$\tau_s$ increases as $r$ increases but the search time is much more
sensitive to $g_0$: the search time changes by about $20\%$ for $g_0
\approx 0$ and by about $0.05\%$ for $g_0 \approx 50$ over the whole
range of $r$. In our model, the search time is not greatly affected by
whether lesions reflect or absorb electrons; what is important is that
the lesions prevent their passage along the DNA.

Figure \ref{fig6}(b) shows that for the range of $\eta$ values
explored, there is a value $0< \sigma^* < 1$ for which the search time
$\tau_{s}$ is minimum.  To understand why there is an optimal
$\sigma^*$, consider the CT mechanism's dependence on the binding
affinity $\sigma$. If $\sigma=1$, enzymes strongly bind onto the
DNA. Even when they absorb electrons, they will re-emit them to stay
adsorbed on the strand. Hence, there is no desorption, no fast
transport through the cytoplasm and acceleration of the search. On the
other hand, if $\sigma=0$, enzymes do not stay on the DNA long enough
to even slide into lesions and the search is correspondingly slow. Our
results show that the search is optimal when the enzyme is
\emph{weakly associated} with the DNA \textit{i.e.} $0<\sigma \ll 1$
and the electron release rate is small compared to the intrinsic
desorption rate. From the data in Tables \ref{tab:constants} and
\ref{tab:constants2}, it seems that real cells do not operate near
this optimal regime.  In Fig. \ref{fig6}(b), we computed most of the
search times using unrealistically large values of $\eta$ to clearly
show the minimum with respect to $\sigma$. For smaller $\eta$ values
we have $\sigma^* \to 0^+$, but the dependence of $\tau_s$ on $\sigma$
does not change qualitatively.  For larger $\eta$ values, enzymes do
not rely on CT to localize to lesions and can find their targets
quickly using diffusive sliding.  In this case, the search is most
efficient if as many enzymes as possible adsorb on the DNA; this
situation is realized by taking $\sigma = 1$ and $\tau_s$
monotonically increases as $\sigma$ gets smaller.

%

Figure \ref{fig7}(a) shows how the search time varies as a function of
1D enzyme diffusivity along the DNA. Notice that the search time for CT
enzymes ($U=1$) is much smaller than that for passive enzymes
($U=0$). Indeed, $\tau_s$ can be reduced by several orders of
magnitude when the effects of CT are included.  If fewer oxoGs are
initially present, the search occurs more quickly.  Consistent with
Fig. \ref{fig6}(a), the search time is extremely sensitive to the
initial number of guanine radicals on the DNA.  For passive enzymes,
$\tau_{s}$ scales as $O(\eta^{-1})$. For CT enzymes, the
$O(\eta^{-1})$ behavior switches to $\tau_{s} = O(\eta^{-1/3})$ for
sufficiently large $\eta$ with the crossover dependent on $g_0$.

%
For finite copy number Fig. \ref{fig7}(b) again shows that the search
time decreases if CT is included, but this time for different flip
rates.  For the large values of $F$ used in Fig. \ref{fig7}, one can
show that the effective diffusion coefficient of the electron density
scales as $1/F$ \cite{FokChou}.  Therefore, as $F$ increases the
electron density dissipates more slowly through the partially
absorbing lesions. A greater density of free electrons implies more
enzyme desorption, more transport through the cytoplasm and faster
search times.  In the $F \to \infty$ limit, we expect the enzymes to
self-desorb independently of the oxoG density. This can be seen from
Eqs. \ref{eqn:fNp} and \ref{eqn:fNm} where the dominant terms on the
right hand side are $\pm F(N_- - N_+)$ and $\sigma R_b/2$. In
principle, as $F\rightarrow \infty$, one can approximate $N_{\pm}$ in
terms of $R_b(t)$ and substitute into $-U(1-\sigma)(N_+ + N_-) Q$ in
Eq. \ref{eqn:fQ} to further reduce the system to only two equations
for $Q(x,t)$ and $R_b(t)$.
%

Although both plots in Fig. \ref{fig7} are for the finite copy number
case, we also performed analogous simulations for the infinite copy
number limit. We found that including the effects of CT by taking
$U=1$ always led to an increase in the search time compared to the
passive case: for fixed $\eta$ and $F$, increasing $U$ always
increased $\tau_s$ regardless of the value of $g_0$.

\section{Conclusions}

Our key finding is that charge-transport (CT) dynamics mediated by
redox reactions can significantly reduce search times of repair
enzymes in real cells where the copy number is finite and the
diffusivity along the DNA is small. In theoretical systems where the
copy number is infinite, CT actually slows down the search.  The
speed-up in finite systems arises because of a spatially dependent
desorption rate. Specifically, the desorption is greater along intact
portions of the DNA but smaller near lesions.  As a result, CT-induced
enzyme-enzyme interactions ``recycle'' enzymes so that they desorb
from lesion-free parts of the DNA and reattach closer to lesion
sites. Our proposed mechanism is illustrated in Fig. 8. A related
mechanism has been implicated in mRNA translation where ribosomes are
recycled, enhancing protein production rates.  \cite{CHOURNA}

%
If we re-dimensionalizing the search times by using an estimated value
of $\kon = 2000$ s$^{-1}$, we find that passive enzymes with
diffusivity $\eta \sim 10^{-10}$ have long search times $\tau_{s}$ of
about 15 minutes (see Fig. \ref{fig7}), comparable to the the life
cycle time of \textit{E. coli}. With the CT mechanism and $g_0 = 28$
initial oxoGs, the search time drops to a few seconds.  For smaller
values of $\eta$, the difference in search times between passive and
CT enzymes becomes greater.  Using $g_{0} = 20$, we calculate
$\tau_{s}$ to be about 30 hours and 2 seconds respectively for passive
and CT enzymes that have diffusivity $\eta \sim 10^{-12}$.  Therefore
for realistic enzyme diffusivities, we think that CT is an
indispensable mechanism that allows enzymes such as MutY to locate
lesions on the DNA in a reasonable amount of time.
%

When the initial number of oxoGs exceeded the enzyme copy number, we
found a large increase in the search time. In this case, search takes
place mostly through slow diffusive sliding along the DNA.  However,
when the copy number (number of potential electron emitters) exceeds
the number of electron absorbers, we find that the search time
decreased drastically, with the search taking place mainly through the
transport of enzymes through the cytoplasm. Therefore, we predict that
the spontaneous generation of electron absorbing defects (such as
oxoG) would significantly slow down the search and conversely, the
presence of other redox-active proteins (such as the transcription
factor SoxR \cite{HidalgoDemple}) would speed up the search. Although
such proteins may not be directly involved in lesion search, they may
be upregulated when the cell is oxidatively stressed, increasing the
population of electron emitters in the system.
The iron-sulfur cluster responsible for CT in MutY is also found in
other repair enzymes like EndoIII \cite{PARIKH}. Hence EndoIII could
also participate in CT, emit electrons to promote the desorption of
MutY and speed up the search.

Recall that classical facilitated diffusion theory
\cite{Berg81,Winter89, vonHippelBerg89} predicts a large reduction in
the search time of proteins providing equal amounts of time are spent
in 1D and 3D. However, most proteins are strongly associated with DNA
so that the speed up is not achieved in practice. The passive enzyme
system considered in this study can be thought of as a suboptimal
search by facilitated diffusion: with $U=0$, a MutY that oxidizes and
binds to the DNA cannot desorb back into the cytoplasm and the protein
spends much more time diffusing in 1D. However, when $U>0$, bound
oxidized MutY can be knocked off the strand by electrons. CT therefore
provides a mechanism for MutY to spend more time in 3D than it
otherwise would. In other words, CT-aided MutY could be one system
where the conditions required for speed up are actually satisfied.
In addition, when MutY binds near lesions, it may diffusively slide
along the DNA into its target: the target size is effectively
increased with the DNA acting like an antenna \cite{Grosberg}.  This
antenna effect is enhanced by enzymes preferentially oxidizing and
adsorbing onto parts of the DNA that are near lesion sites.

Extensions to our model may include spatial gradients in the bulk
enzyme concentration, more careful treatment of electron dynamics, and
adding fluctuations in copy number. Nonetheless, our simple
deterministic model describes mechanisms and yields results
qualitatively consistent with findings in \cite{BARTON,BARTON2}.

\bibliography{broad_continuum13.bbl}

\newpage

\section*{Figure Captions}

\noindent Figure 1: (a) Target search on prokaryotic DNA, which is
tightly coiled up into a nucleoid. Proteins in the bulk can diffuse to
the DNA through the cytoplasm to locate their targets.  (b) Searching
proteins (hexagons) locate targets (diamonds) by sliding
along DNA, punctuated by attachment and detachment.  $\Omega$
represents the cell volume while $\Omega'$ represents all points in
the vicinity of the nucleoid.  Enzymes within $\Omega'$ can engage in
direct adsorption onto the DNA.

\vspace{8mm}

\noindent Figure 2: Charge Transport (CT) mechanism proposed in
\cite{BARTON,BARTON2,Boal}. (a) A repair enzyme (in solution) is in the
$2+$ state and adsorbs onto the DNA.  Its iron-sulfur cluster oxidizes
in the process, releasing an electron along the DNA. A repair enzyme
(already adsorbed on the DNA) is in the $3+$ state and accepts an incoming
electron. Its iron-sulfur cluster reduces and the enzyme desorbs.
(b) Guanine radicals (``oxoGs'') can absorb free electrons
on the DNA. These radicals are annihilated upon absorbing an electron.
(c) Lesions can partially reflect and absorb electrons.

\vspace{8mm}

\noindent Figure 3: (a) Summary of the CT model, described by
Eqs. \ref{eqn:CT1}-\ref{eqn:CT6}.  Bulk enzymes, with density
$R_b(t)$, can attach to the DNA and oxidize to release rightward and
leftward electrons with densities $N_+(x,t)$ and $N_-(x,t)$
respectively.  Guanine radicals with density $g(x,t)$ act as electron
absorbers.  Upon adsorption, oxidized enzymes with density $Q(x,t)$
are formed with $R_a(x,t)$ as a transient, intermediate quantity.
Fixed lesions are located at $x=0$ and $x=L$.
(b) Redox reaction diagram for the MutY repair enzyme.
MutY$^{2+}$ in solution is represented by $R_b(t)$,
MutY$^{2+}$ adsorbed onto DNA is represented by $R_a$ and
MutY$^{3+}$ adsorbed onto DNA is represented by $Q$.

\vspace{8mm}

\noindent Figure 4: Density profiles for enzyme, guanine, and
electrons on DNA in a finite enzyme copy number system ($R_b(0) = n_0
= 30$) at time (a) $t=2$ and (b) $t=5$. Dashed lines show density
profiles of passive enzymes in which the CT mechanism is absent.
Parameters used were $\eta=10^{-10}, \sigma=0.9, F=10^5$, and
$g_{0}=28$.

\vspace{8mm}

\noindent Figure 5: Enzyme profiles and currents for infinite ((a) and
(c)) and finite ((b) and (d)) copy numbers. In each figure, the
profile or current is plotted for passive (dashed) enzymes where
$U=0$ and CT (solid) enzymes where $U=1$. 
Insets show the large gradients in enzyme density
within a thin boundary layer near the lesions.  Parameters used were
$\sigma = 0.9$, $\eta=10^{-10}$, $g_0=28$, $r=0.5$, $F=10^5$ and
$n_{0}=30$.

\vspace{8mm}

\noindent Figure 6: Search time $\tau_{s}$ for Charge-Transport
Enzymes, for copy number $n_0=30$, electron flip rate $F=10^5$ and
electron speed $U=1$. The actual search time in seconds can be
recovered by dividing by $\kon$, whose value is estimated in Table
\ref{tab:constants}.
(a) Search time as a function of initial guanine density $g_0$ and
lesion electron reflectivity $r$.  Parameters used were $\sigma=0.9$
and $\eta=10^{-10}$.
(b) Search time $\tau_{s}$ as a function of enzyme binding affinity
$\sigma$ and enzyme diffusivity along DNA, $\eta$. Parameters used
were $g_0=30$ and $r=0.5$.

\vspace{8mm}

\noindent Figure 7: (a) Search time $\tau_{s}$ of passive enzymes
($U=0$) compared with CT enzymes ($U>0$) as a function of enzyme
diffusivity $\eta$ for various initial guanine densities
$g_0$. Parameters used were $\sigma=0.9$, $r = 0.5$, $F=10^5$ and
$n_0=30$.
(b) Search time of passive enzymes compared with CT enzymes for
different electron flip rates $F$. Parameters used were $\sigma=0.9$,
$r = 0.5$, $\eta = 10^{-10}$ and $n_0=30$. For both plots, the actual
search time in seconds can be recovered by dividing by $\kon$, whose
value is estimated in Table \ref{tab:constants}.

\newpage


\vspace{8mm}

\noindent Figure 8: Recyling of enzymes via the Charge-Transport
mechanism. In a finite copy number system, the mechanism increases the
enzyme desorption rate for intact portions of DNA but decreases it
near lesions. Therefore on average, enzymes are ``recycled'' to lesion
sites by 3D transport through the cytoplasm. In many cell systems,
this method of finding lesions is faster than a 1D search by diffusive
sliding.

\newpage
\section*{Figures}

\begin{figure}
\begin{center}
\includegraphics[width=4.5in]{Fig4.eps}
\end{center}
\caption{}
\label{fig4}
\end{figure}

\begin{figure}
\begin{center}
\includegraphics[width=4.5in]{Fig5.eps}
\end{center}
\caption{}
\label{fig5}
\end{figure}

\begin{figure}[htbp]
\begin{center}
\includegraphics[width=4.5in]{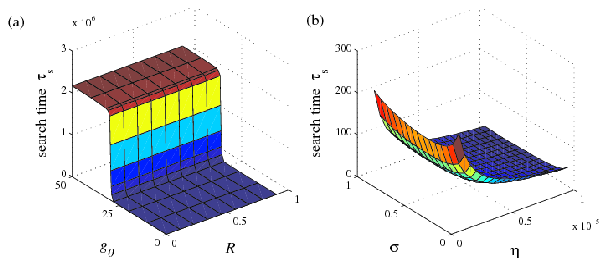}
\end{center}
\caption{}
%
%
%
\label{fig6}
\end{figure}

\begin{figure}[htbp]
\begin{center}
\includegraphics[width=4.5in]{Fig7.eps}
\end{center}
\caption{}
\label{fig7}
\end{figure}

%

\begin{figure}[htbp]
\begin{center}
\includegraphics[width=3.6in]{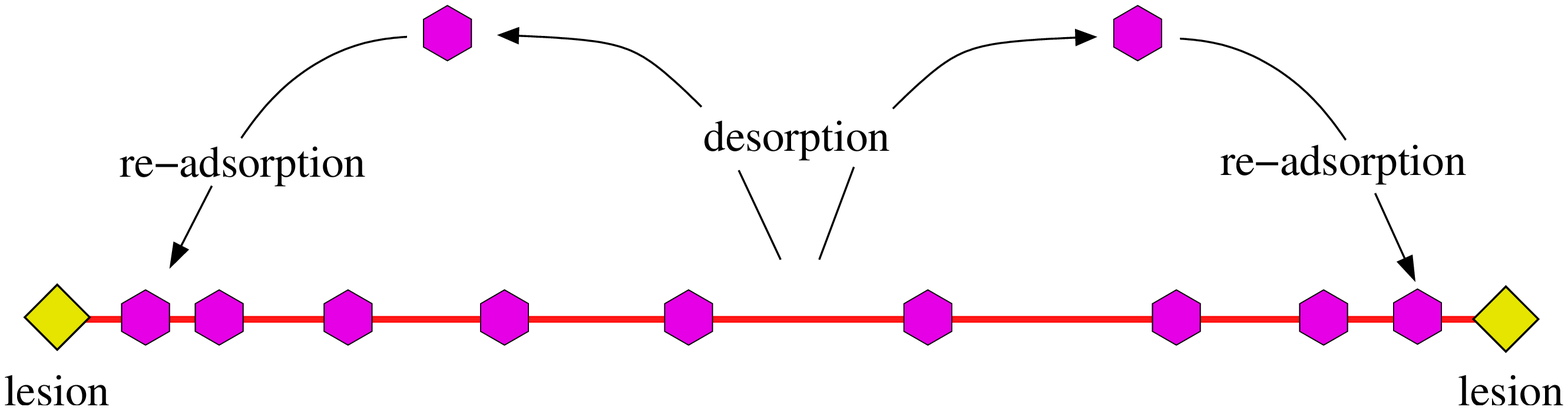}
\end{center}
\caption{}
\label{fig8}
\end{figure}

\end{document}